\documentclass[%
aip,jcp,amsmath,amssymb, reprint, twocolumn,
]{revtex4-2}
\usepackage{amsmath}
\usepackage{mathtools}
\usepackage{float}
\usepackage{amsfonts}
\usepackage{amssymb}
\usepackage{dcolumn} 
\usepackage{array}
\newcolumntype{P}[1]{>{\centering\arraybackslash}p{#1}}
\newcolumntype{M}[1]{>{\centering\arraybackslash}m{#1}}
\newcolumntype{C}[1]{>{\centering\arraybackslwash}p{#1}}
\usepackage{float}
\usepackage{psfrag}
\usepackage{tabularx}
\usepackage{stackengine}
\usepackage{amssymb}
\usepackage{mathtools}  
\usepackage{xfrac} 
\usepackage[T1]{fontenc}
\usepackage{graphicx}
\usepackage{diagbox}
\usepackage[caption=false]{subfig}
\usepackage{tikz}
\usetikzlibrary{positioning}
\usetikzlibrary{arrows}
\usetikzlibrary{trees}
\usepackage{amssymb}
\usetikzlibrary{decorations.pathmorphing}
\usetikzlibrary{decorations.markings}
\usetikzlibrary{automata,positioning}
\usepackage{braket}
\usepackage{hyperref}
\usepackage{rotating}
\usepackage{adjustbox}
\usepackage{ragged2e}
\usepackage{simplewick}
\usepackage{simpler-wick}

\usepackage[]{lineno}

\usepackage{csquotes}
\usepackage{physics,amsmath}
\usepackage{xcolor}
\usepackage{fancyhdr}
\UseRawInputEncoding
\usepackage[normalem]{ulem}

\usepackage{scalerel}

\begin{document}

\author{Chinmay Shrikhande}
\altaffiliation{Authors contributed equally to this work}
\affiliation{ Department of Chemistry,  \\ Indian Institute of Technology Bombay, \\ Powai, Mumbai 400076, India}

\author{Sonaldeep Halder}
\altaffiliation{Authors contributed equally to this work}
\affiliation{ Department of Chemistry,  \\ Indian Institute of Technology Bombay, \\ Powai, Mumbai 400076, India}

\author{Rahul Maitra}
\email{rmaitra@chem.iitb.ac.in}
\affiliation{ Department of Chemistry,  \\ Indian Institute of Technology Bombay, \\ Powai, Mumbai 400076, India}
\affiliation{Centre of Excellence in Quantum Information, Computing, Science \& Technology, \\ Indian Institute of Technology Bombay, \\ Powai, Mumbai 400076, India}

\title{Development of Zero-Noise Extrapolated Projection Based Quantum Algorithm for Accurate Evaluation of Molecular Energetics in Noisy Quantum Devices}

\begin{abstract}
The recently developed Projective Quantum Eigensolver (PQE) offers an elegant procedure to evaluate the ground state energies of molecular systems on quantum computers. However, the noise in available quantum hardware can result in significant errors in computed outcomes, limiting the realization of quantum advantage. Although PQE comes equipped with some degree of inherent noise resilience, any practical implementation with apposite accuracy would require additional routines to suppress the errors further. In this work, we propose a way to enhance the efficiency of PQE by developing an optimal framework for introducing Zero Noise Extrapolation (ZNE) in the nonlinear iterative procedure that outlines the PQE; leading to the formulation of ZNE-PQE. For this method, we perform a detailed analysis of how various components involved in it affect the accuracy and efficiency of the reciprocated energy convergence trajectory. Moreover, we investigate the reasons behind the improvements observed in ZNE-PQE over conventional PQE by performing a comparative analysis of their residue norm landscape. This approach is expected to facilitate practical applications of quantum computing in fields related to molecular sciences, where it is essential to determine molecular energies accurately.

\end{abstract}

\maketitle
\section{Introduction}
The prospect of efficient simulation of molecular systems in
the quantum computing framework has attracted an insurmountable amount of
research interest. Various quantum algorithms have been 
developed which aim to obtain the ground and excited state energies of 
molecular systems. Unlike the classical methods, these algorithms leverage
the superposition, interference, and entanglement principles provided by quantum hardware
to manipulate complex molecular wavefunctions in a tractable way\cite{wecker2014gate, poulin2014trotter, kassal2011simulating, reiher2017elucidating}. But in order to surpass their classical counterparts, these algorithms require fault-tolerant quantum hardware. However, the currently available quantum devices suffer from various limitations in the form of poor gate fidelity, state preparation and measurement errors (SPAM), crosstalk, and limited coherence\cite{knill1998resilient, ladd2010quantum}. In order to achieve practical utility in implementing any quantum algorithm in Noisy Intermediate-Scale Quantum (NISQ)\cite{preskill2018quantum} devices, one requires  precise calibrations and the additional use of error-suppressing techniques.

Quantum error correction (QEC)\cite{shor1995scheme, calderbank1996good, steane1996error} provides the pathway towards scalable fault-tolerant quantum 
computation where each logical qubit is mapped to multiple physical qubits. These techniques involve adaptive corrections to eliminate errors that are produced during the execution of any quantum circuit. But to construct a system with sufficient logical qubits that can encode molecules or atoms, one would require a quantum device with an exorbitant number of actual qubits.
For example, the \textit{surface code}\cite{dennis2002topological, kitaev2003fault, raussendorf2007fault} requires \textit{O}($d^2$) 
physical qubits to simulate a single logical qubit, d being associated to 
the number of errors the method can correct. As the availability of a large number of physical qubits is beyond the scope of current NISQ hardware, the application of QEC becomes practically constrained. Moreover, most current quantum hardware does not support adaptive rectification. Apart from these limitations, a variety of QEC methods rely on the exact characterization of the noise associated with the quantum hardware. While procedures such as Gate set tomography, process tomography, and randomized benchmarking\cite{chow2009randomized, blume2017demonstration, chuang1997prescription, poyatos1997complete} have been developed for this characterization, they either do not scale properly to larger quantum systems or only partially describe the noise\cite{dankert2009exact, emerson2005scalable, emerson2007symmetrized, da2011practical, flammia2011direct}. These limiting factors make it unattainable to incorporate QEC in problems that require sufficiently large quantum systems and deep circuit implementation.

Meanwhile, a suite of \textit{error mitigation}\cite{temme2017error, kandala2019error, endo2018practical, bonet2018low, li2017efficient, huggins2021virtual, song2019quantum, zhang2020error, koczor2021exponential} techniques promises to increase the practical applicability of NISQ devices without the need for additional quantum resources (such as qubits). These methods rely on the outcome from an ensemble of measurement data that 
are further post-processed outside the quantum architecture to estimate the outcome from ideal quantum circuits. Some of them are even designed to function without the need for accurate noise characterization of the hardware. In the NISQ era, these methods provide the most practical and efficient solution to the problem of noise. It is theoretically possible to obtain an upper or lower bound to their efficacies as done by R. Takagi et al\cite{takagi2022fundamental}. However, the actual performance (and feasibility) of any error-mitigation technique varies from case to case and is highly dependent on the platform to which it is applied. In this manuscript, our primary focus is on the algorithms pertaining to electronic structure theory. Given the depth of the circuits utilized during the implementation of different molecular simulation algorithms and the limitations associated with the exact characterization of the hardware noise, Zero Noise Extrapolation (ZNE)\cite{temme2017error, kandala2019error, giurgica2020digital, he2020zero} offers the most suitable solution. In ZNE, the estimate of a noiseless expectation 
value is obtained by post-processing the results from a given number of experiments, each performed at different augmented noise strengths. The post-process step involves extrapolating the curated measurement results to obtain the zero noise value. It can be efficiently used to mitigate collective gate errors without the specific characterization of the employed NISQ hardware. 

In this work, we develop the optimal architecture of zero noise extrapolated projective quantum 
eigensolver (or ZNE-PQE)\cite{stair2021simulating}, an iterative algorithm aimed at determining the accurate ground 
state energy of molecular systems in the noisy quantum hardware. The conventional PQE provides a description for simulating many-fermion systems. It relies on the 
construction of a parameterized trial wavefunction using the dUCC ansatz 
on quantum hardware. The optimum parameters correspond to the solution of 
a set of non-linear equations fabricated by projecting the Schr\"{o}dinger 
equation with a set of many-body basis. The quasi-Newton-type iterative 
procedure is utilized to arrive at the numerical solution to these equations whose convergence demarcates the ground state energy. Although PQE is 
shown to have a high noise resilience under a stochastic noise model, ZNE-PQE provides a promising approach towards improvement in its practical applicability and 
efficacy for various chemical systems in real quantum architectures. Moreover, it can be clubbed with many of its cost-efficient variants to lower the overall measurement overhead\cite{sonaldeep2023, magoulas2023cnot, magoulas2023linear}. During the implementation of our method, we perform an in-depth analysis of 
various components involved in ZNE-PQE and how they affect the iterative energy 
trajectory under different noise characteristics. Furthermore, we compare the residue norm landscape of ZNE-PQE with the conventional PQE to ascertain the underlying reasons that lead to superior performance in the ZNE-PQE.

The theory has been divided into two parts - in section II A, we present the fundamental concepts associated with conventional PQE and illustrate its implementation on IBM's superconducting quantum devices viz \textit{ibmq\_manila} and \textit{ibmq\_belem}. Within this context, we provide a concise overview of the prevailing sources of noise, which produces errors during computational processes. Furthermore, we expound upon the manner in which these sources of noise are effectively modeled by means of quantum error channels. In section II B, we start by describing the general structure of any error mitigation process along with the metrics that highlight its efficacy. We move on to the particular case of ZNE and provide the theoretical upper bound to its generated estimates. Next, we formulate the construction of ZNE-PQE, which aims to produce accurate ground state energies within the domain of noisy quantum hardware. We outline the various modules that collectively constitute ZNE-PQE and explore the varieties associated with them.  

\section{Theory}

\begin{figure*}[!ht]
    \centering
\includegraphics[width=\textwidth]{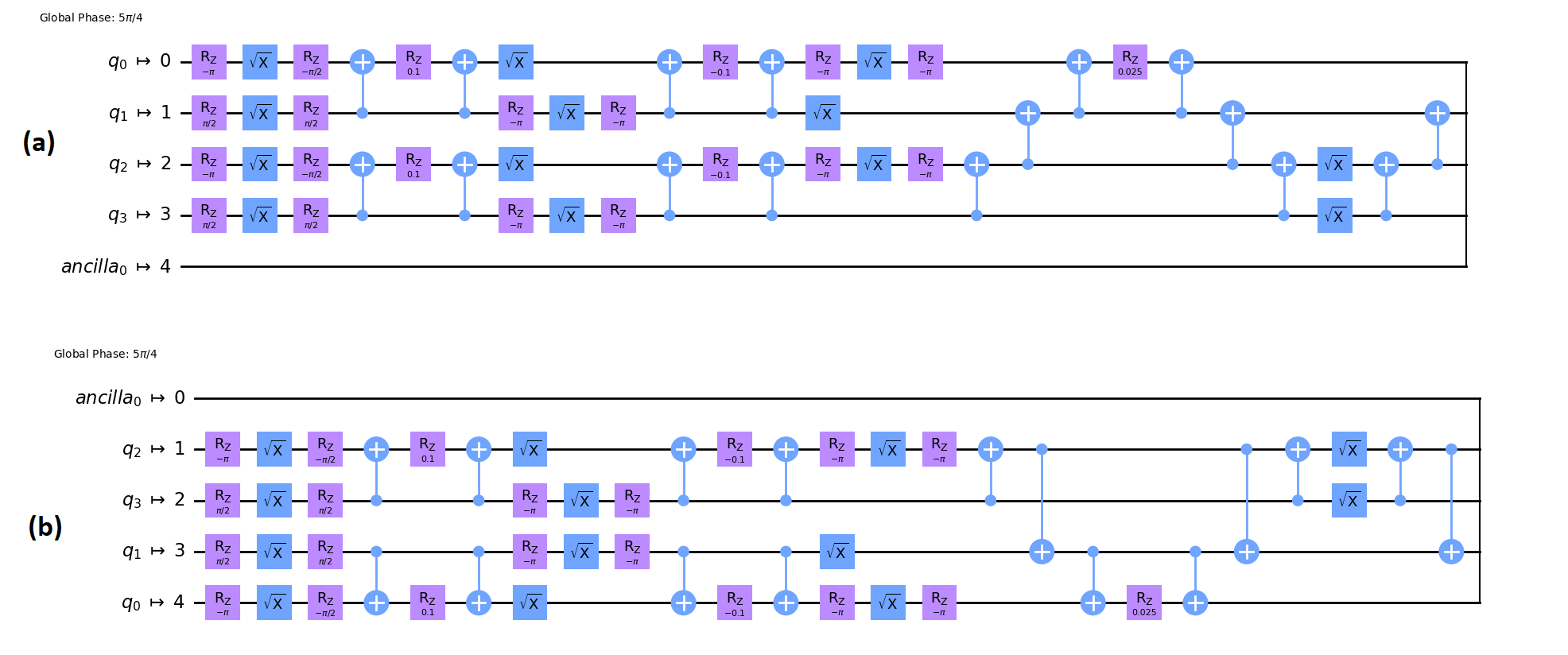}
\caption{A portion of the circuit implemented to evaluate Term 3 for $H_2$ on (a)ibmq\_manila and (b)ibmq\_belem. The ansatz parameters were set at an arbitrary value of [0.1, 0.1, 0.1]. These two quantum devices are five qubits each. This figure highlights which qubits are employed during the simulation of a four-qubit chemical system (such as $H_2$) and which are left out (represented as ancilla).}
    \label{circuits}
\end{figure*}

\subsection{Projective Quantum Eigensolver: Theoretical Aspects and Circuit Depiction}
In PQE, an approximate trial ground state($\ket{\Psi (\theta)}$) is obtained by the action of a parameterized unitary ansatz
($\hat{U}(\theta)$) on a reference state ($\ket{\Phi_o}$),

\begin{equation} \label{ansatz_action}
   \ket{\Psi (\boldsymbol{\theta)}} = \hat{U}(\boldsymbol{\theta})\ket{\Phi_o}.
\end{equation}

Inserting this in the Schr\"{o}dinger equation and subsequently left multiplying by $\hat{U}^{\dagger}(\theta)$, we get

\begin{equation}
   \hat{U}^{\dagger}(\theta)\hat{H} \hat{U}(\theta)\ket{\Phi_o} = \mathcal{E}\ket{\Phi_o}.
\end{equation}

Now, projecting with the reference state ($\ket{\Phi_o}$) gives the energy equation for PQE, whereas projecting with the complete orthonormal many-body basis, excluding the reference state$\{\ket{\Phi_{\mu \ne o}}\}$, gives the residual conditions which are used to determine the optimal parameters($\boldsymbol{\theta}$),

\begin{equation} \label{3}
   \bra{\Phi_o}\hat{U}^{\dagger}(\theta)\hat{H} \hat{U}(\theta)\ket{\Phi_o} = \mathcal{E}_{PQE}(\theta) \\
\end{equation}

\begin{equation} \label{4}
       r_{\mu}(\theta) = \bra{\Phi_{\mu}}\hat{U}^{\dagger}(\theta)\hat{H} \hat{U}(\theta)\ket{\Phi_o} = 0
\end{equation}

Here, $\mu$ runs over the entire many-body basis(except the reference, i.e., $\mu \ne o$). For cases where the number of parameters is less than the number of many-body basis, the residual condition(Eq. \eqref{4}) can only be imposed on a subset of these residuals ($r_{\mu}$). This results in uncertainty in the PQE energy ($\mathcal{E}_{PQE}$) given by the Gershgorin circle theorem; according to which the energy difference ($|\mathcal{E}-\mathcal{E}_{PQE}|$) is bound by a circle of radius $\ell$ as:

\begin{equation}
    |\mathcal{E}-\mathcal{E}_{PQE}| \leq \ell
\end{equation}

While the PQE equations are general, our study focuses on disentangled Unitary Coupled Cluster (dUCC) ansatz\cite{evangelista2019exact}. It is characterized by a pool of ordered, non-commuting, anti-hermitian particle-hole operators ($\{\hat{\kappa}\}$) with the reference state taken to be the single Hartree Fock (HF) determinant $\ket{\Phi_o} = \ket{\chi_i\chi_j...}$, where $\chi's$ are the spin-orbitals. The ansatz can be written as,

\begin{equation}\label{6}
    \hat{U}(\theta) = \prod_{\mu} e^{\theta_{\mu}\hat{\kappa}_{\mu}}
\end{equation}
 
\begin{equation}
    \hat{\kappa}_{\mu}=\hat{\tau}_{\mu}-\hat{\tau}_{\mu}^\dagger
\end{equation}
   
\begin{equation}
    \hat{\tau}_{\mu}= \hat{a}^{\dagger}_{a}\hat{a}^{\dagger}_{b}....\hat{a}_{j}\hat{a}_{i}
\end{equation}

In the above equations, $\mu$ represents a multi-index particle-hole excitation structure as defined by the string of creation($\hat{a}^{\dagger}$) and annihilation($\hat{a}$) operators. Indices \{i,j...\} denote occupied spin-orbitals in the HF state, and \{a,b,...\} denote the unoccupied spin-orbitals. $\hat{\kappa}_{\mu}$ acts on the reference state to generate a many-body basis,

\begin{equation}
    \hat{\kappa}_{\mu}\ket{\Phi_o} = \ket{\Phi_{\mu}}
\end{equation}

In practical applications, dUCC ansatz is constructed using only a subset of the total possible excitation operators($\hat{\kappa}_{\mu}$), and the residual condition(Eq. \eqref{4}) is implied over the corresponding basis.
Eqs. \eqref{3} and \eqref{4} represent a set of non-linear equations in $\{\theta_{\mu}\}$ which can be solved numerically via quasi-newton iterative scheme:
\begin{equation} \label{iter}
    \theta_{\mu}^{(k+1)} = \theta_{\mu}^{(k)} + \frac{r_{\mu}^{(k)}}{\Delta_{\mu}}
\end{equation}

Here, $k$ represents an iteration level and $r_{\mu}$ is the residue as defined 
by Eq. \eqref{4}, which is calculated on a quantum computer. The 
denominator $\Delta_{\mu}$ depends on the excitation structure of $\mu$ and is 
given as $\epsilon_{i} + \epsilon_{j} + ...... - \epsilon_{a} - \epsilon_{
b} 
- ....$; $\epsilon_p$ is the HF energy of the $p^{th}$ spin-
orbital and $\{i,j,..\}$ are the orbital indices in HF 
determinant($\ket{\Phi_o}$) which has been replaced by orbitals of indices $\{a,b,...\}$ to form the excitation structure of $\mu$. Usually, the iterative scheme is deemed converged when the $2-norm$ of the residue vector 
($\| r\|$) goes below a predefined threshold (usually taken to be a small 
value like $10^{-05}$ or $10^{-06}$). The expression,  $r_{\mu} = 
\bra{\Phi_{\mu}}\Bar{H}\ket{\Phi_o}$, where $\Bar{H}=\hat{U}^{\dagger}
(\theta)\hat{H}\hat{U}(\theta))$, can be  further simplified if we consider the action 
of $e^{\omega\hat{\kappa}_{\mu}}$ on the reference state,

\begin{equation} \label{11}
\begin{split}
\ket{\Omega_{\mu}(\omega)} & = e^{\omega \hat{\kappa}_{\mu}}\ket{\Phi_o}\\
&=cos(\omega)\ket{\Phi_o} + sin(\omega)\ket{\Phi_{\mu}}
\end{split}
\end{equation}

The last equality of Eq. \eqref{11} holds true because $\hat{\kappa}_{\mu}$ follows the relation: $\hat{\kappa}_{\mu}\ket{\Phi_o} = \ket{\Phi_{\mu}}$ and $\hat{\kappa}^{2}_{\mu}\ket{\Phi_o} = -\ket{\Phi_{\mu}}$. Setting $\omega = \frac{\pi}{4}$ and rearranging Eq. \eqref{11}, we get

\begin{equation} \label{12}
    r_{\mu} = D_{\Omega_{\mu}} -\frac{D_{\mu}}{2} - \frac{D_o}{2}
\end{equation}
Here,$D_{\Omega_{\mu}}=\bra{\Omega_{\mu}(\frac{\pi}{4})}\Bar{H}\ket{\Omega_{\mu}(\frac{\pi}{4})}$, $D_{\mu} = \bra{\Phi_{\mu}}\Bar{H}\ket{\Phi_{\mu}}$ and $D_o = \bra{\Phi_o}\Bar{H}\ket{\Phi_o}$. Eq. \eqref{12} enables us to calculate the residues only by measuring the diagonal terms $\{D\}$ on a quantum computer. In the rest of the article, $D_o$ is often referred to as Term 3. The iterative prescription described by Eq. \eqref{iter} in conjunction with Eq. \eqref{12} gives a complete recipe for evaluating the ground state energies of molecular systems on quantum hardware. Fig. \ref{circuits} shows the circuit implemented to determine Term 3 for $H_2$ at an arbitrary set of parameters ($\{\theta\}=0.1$) on two real IBM quantum devices viz \textit{ibmq\_manila} and \textit{ibmq\_belem}. Both of these are five qubit systems each with their properties and connectivities provided in the Supplementary Material. 

During the evaluation of the terms delineated by Eq. \eqref{12}, the unwanted interaction of the supposedly closed quantum system with the environment  inevitably introduces errors in the measured values. The dominant channels used to describe these interactions include the depolarizing channel and thermal relaxation error channel (which represents the combined amplitude and phase damping). Additionally, the inaccurate process of qubit state acquisition during measurements results in readout errors. The depolarizing channel represents the transformation of a quantum state into a linear combination of itself and the maximally mixed state ($I/d$, $d$ is the dimension of the Hilbert Space formed by the qubit states). This channel suitably accounts for the depolarization of the qubits upon their interaction with the environment during the implementation of a quantum gate. Amplitude damping describes the comprehensive scenario where a system exchanges energy with the environment. During such processes, the qubit is driven to either $\ket{0}$ or $\ket{1}$ state. On the other hand, phase damping describes the system's loss of quantum information or coherence without any loss in energy. Readout errors arise due to incorrect recording of the measured qubit state. It is described by $p(X|Y)$, which is the probability that the noisy measurement is recorded as the qubit-state $X$, whereas the true measurement outcome was $Y$. A detailed description entailing the noise channels that have been discussed above can be found in the works of Georgopoulos \textit{et. al}\cite{georgopoulos2021modeling}.

While conventional PQE is shown to be inherently noise resilient,  the incorporation of additional layers of error mitigation techniques holds tremendous potential for significantly enhancing its accuracy. This augmentation, in turn, would enable achieving quantum advantage using NISQ hardware for practical, real-life applications in the realm of molecular systems. In the next section, we provide a theoretical description of these mitigation techniques, ZNE in particular, while outlining their theoretical limits for producing an estimate of the ideal value. We proceed to establish the foundational framework essential for the seamless integration of ZNE within the broader PQE paradigm. This gives rise to ZNE-PQE, an iterative scheme that promises to obtain accurate ground state energies of molecular systems in NISQ devices.

\subsection{General bounds of ZNE and Development of ZNE-PQE}
As described by R. Takagi et al.\cite{takagi2022fundamental}, any general error mitigation technique can be described by a physical process $\mathcal{P}$ which takes in an ensemble of $N$ distorted states $\{\epsilon_i(\rho)\}_{i=1}^N$ generated via the effective noise channels $\{\epsilon_i\}_{i=1}^N$ and gives out the estimate random variable $E_A$ of $Tr(A \rho)$, $\rho$ being the ideal state and $A$ representing the operator for a given observable. Any general method for error mitigation can be represented as a $(Q, K)$-mitigation protocol that involves $Q$ inputs and $K$ experiments. Each experiment is associated with a Positive Operator-Valued Measure (POVM) denoted as $\{M_{i^{(K)}}^{(K)}\}$, which forms the foundational basis for the quantum-classical channel. This channel subsequently generates an ensemble of outcomes denoted as $ \textbf{i}= \{i^{(1)},\dots,i^{(K)}\}$, thereby encapsulating the collective results obtained from the aforementioned experiments. The classical-classical channel $\hat{e}_A$  processes the ensemble of outputs and generates the estimated value to an ideal expectation value.

\begin{equation}
    \langle E_A \rangle = \sum_{\textbf{i}}p_{\textbf{i}}e_A(\textbf{i}) = Tr(A\rho) + b_A(\rho)
\end{equation}
 Here, the sum over $\textbf{i}$ takes into account the spread in the estimation. The entire process of the $(Q, K)$-mitigation method can be thought of as a tandem application of the two above-mentioned channels.
 
The performance of a given error mitigating method is captured by the bias, which is a measure of the absolute minimum error with which it can estimate $\Tr(A\rho)$ i.e.
\begin{equation}
    b_A(\rho) = \langle E_A \rangle-Tr(A\rho).
\end{equation}

As any error mitigation technique is independent of the ideal state $\rho$ and the observable $A$, the maximum bias can be written as:
\begin{equation}
    b_{max}= \underset{-\frac{I}{2} \leq A \leq \frac{I}{2}}{\text{max}} \underset{\rho}{\text{max}} |\langle E_A \rangle - Tr(A\rho)|
\end{equation}

Here, the limits to $A$ are general, as any operator can be put to these bounds by shifting and re-scaling them. Another important aspect that determines the efficacy of a mitigation method is the spread in the estimates, $\Delta e_A$. This metric indicates the difference between the highest and lowest achievable values for $E_A$. The maximum spread for any arbitrary $A$ can be represented as:
\begin{equation}
    \Delta e_{max} = \underset{-\frac{I}{2} \leq A \leq \frac{I}{2}}{\text{max}} \Delta e_A
\end{equation}

The smaller the spread, the lower the variance, which in turn translates to concentrated estimates for the mitigation output.

Apart from the performance of a given mitigation method, it is essential to assess the feasibility of the same for a given quantum algorithm. Among the various available methods, ZNE provides a promising platform for 
accessing accuracy beyond what is offered by the native noisy quantum hardware.   
The method can be employed to mitigate gate-based errors without an exact 
comprehension of the associated noise channels and does not require 
additional qubits. These properties make it highly suitable to be used with 
various quantum algorithms that deal with the domain of molecular energetics, which involves deeper circuits with many one and two-qubit gates.
ZNE employs an arrangement comprising an ensemble of noise channels denoted as $\{\mathcal{N}_{\lambda}\}$, where each noise channel is associated with a particular noise strength represented by the parameter $\lambda$. Although the precise characterization of these noise channels remains ambiguous, the underlying methodology assumes the availability of means to enhance the noise strength, such that $\lambda \geq \lambda_o$. The value of $\lambda_o$ represents the native noise strength present in a given circuit. The essence of ZNE is that by learning the relationship between $\lambda$ and the expectation value of an observable at noise level $\mathcal{N}_{\lambda}$, we can extrapolate it to its value at $\lambda=0$. Taking the case of $R^{th}$ order Richardson extrapolation, it can be established that the extrapolated estimate can be written as a linear combination of $R+1$ expectation measurements obtained at different nodes ($\{C_r\}_{r=0}^R$):

\begin{equation}
    \sum_{r=0}^R \gamma_r Tr(A\mathcal{N}_{C_r\lambda_o}(\rho)) = Tr(A\rho) + b_A(\rho)
\end{equation}
where,
\begin{equation}
    \sum_{r=0}^R \gamma_r =1
\end{equation}

\begin{align}
    \sum_{r=0}^R \gamma_r C_r^q = 0 \quad q = 1, \dots , R
\end{align}

Now, let $\{a^K\}_{K=1}^{R+1}$ denote the set of outputs obtained through the POVM measurements $\{M_{a^K}^K\}_{K=1}^{R+1}$ corresponding to the projectors in the spectral decomposition of a general operator $A$. The estimator can be given as:

\begin{equation}
    e_A(\{a^K\}_{K=1}^{R+1}) = \sum_{K=1}^{R+1} \gamma_{K-1}a^{(K)}
\end{equation}

Since, the operator $A$ is constrained to $-\frac{I}{2} \leq A \leq \frac{I}{2}$, this puts the bound $-\frac{1}{2} \leq a \leq \frac{1}{2}$ on the measurement outcome. This, in turn, generates the bound to the maximum and minimum possible value of the estimator function:

\begin{equation}
    e_{A, max} \leq \frac{1}{2}\sum_{r=0}^{R}|\gamma_r|
\end{equation}

Similarly, 

\begin{equation}
    e_{A, min} \geq -\frac{1}{2}\sum_{r=0}^{R}|\gamma_r|
\end{equation}

Thus, the maximum spread can be represented as:

\begin{equation}
    \Delta e_{A,max} \leq e_{A, max} - e_{A, min}= \sum_{r=0}^{R}|\gamma_r|
\end{equation}
The exact spread will depend on the quantum algorithm to which ZNE is applied. The molecular simulations using PQE provide a stimulating opportunity since the working equations (Eqs.\eqref{iter}, \eqref{12}) deal with the measurement of diagonal terms.
These diagonal values can be parameterized with the system noise level $\lambda$. For convenience, we rescale the noise strengths by setting $\lambda_o$ to 1. If $\mathcal{N}_{\lambda}(\rho)$ is the output state after the action of unitary $\hat{U}(\theta)$ on HF state (Eq. \eqref{ansatz_action}) for the system where noise is scaled up to $\lambda$, the diagonal terms (Eq. \eqref{12}) can be given as:
\begin{equation}
    D(\lambda) = Tr(H\mathcal{N}_{\lambda}(\rho))
\end{equation}
where $\lambda = 0$ maps to the noiseless expectation value.
By sequentially evaluating the expectation, $D(\lambda)$, at enhanced noise scales and extrapolating it to obtain the value of $D(\lambda=0)$, gives rise to the formulation of ZNE-PQE. It promises to achieve accurate molecular energies without spending any additional qubits. Although, it does require a large measurement overhead. The efficacy of the ZNE-PQE is largely dependent on two factors:
\begin{itemize}
    \item the methodology used to augment the noise strength within a NISQ device to get the assembly of $\{D(\lambda)\}_{\lambda}$.
    \item the extrapolating method utilized to estimate the value beyond the range of available values, i.e., $D(0)$.
\end{itemize}

\begin{figure*}[!ht]
    \centering
\includegraphics[width=\textwidth]{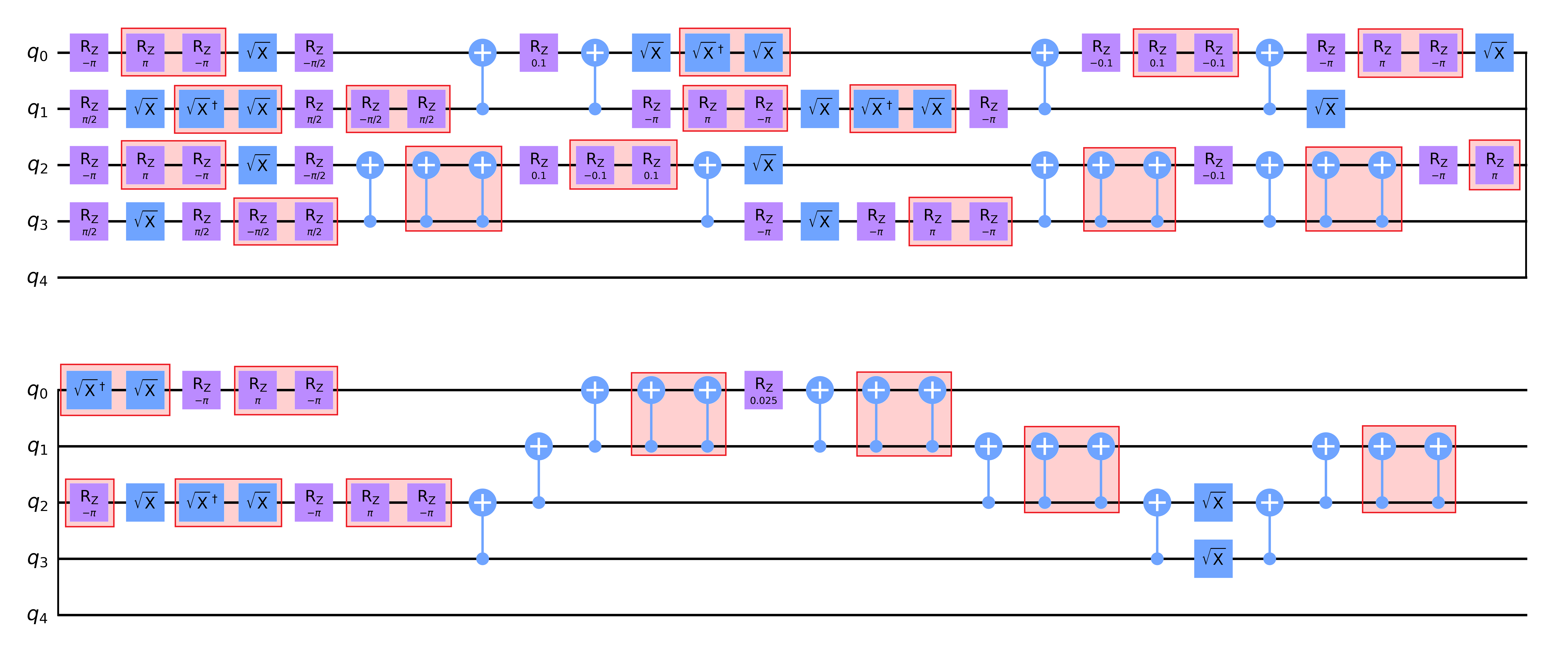}
\caption{Local unitary folding on a portion of the circuit involved in the determination of $D_o(\lambda=2)$ on $ibmq\_manila$ for $H_2$ molecular system. In this illustrative example, the gates are folded randomly at positions marked by red boxes for the noise scale factor $\lambda$ of 2. }
    \label{manila_scaled}
\end{figure*}

\subsubsection{Noise Scaling Methods}
The effective noise in the system can be scaled up using time-scaling 
methods on the quantum processors. This is executed by re-calibrating the control pulse for every gate, which leads to the same unitary evolution but requires a longer time. Another direct method to scale the noise spectrum is by manipulating the master clock of the system, which 
requires precise noise characterization. These methods are inefficient 
due to their resource-intensive nature, requiring low-level access to 
the quantum hardware for successful implementation and difficulty in precise characterization. 

An alternative approach to scale noise is the gate-based noise injection 
method, where rather than increasing the time for which the gate is applied,
the effective number of gates is increased. It is done in a 
way that does not change the logical action of the entire circuit, but the 
increased circuit depth results in a scaled-up noise. This process can be 
considered as the reverse of \textit{circuit optimization}, which involves 
strategic reduction of circuit depths. The advantage of such an approach is that it does 
not require the characteristic knowledge of noise channels tied with a 
quantum processor.

The state preparations involved in the ZNE-PQE framework are based on the unitary ansatz ($\hat{U}(\theta)$) realized in terms of one and two-qubit gates. The scaled circuit can be constructed as follows: 
\begin{equation}
    U \rightarrow U(U^{\dagger}U)^n
\end{equation}
where $n$ is a positive integer. Although the 
application of the transformed operator generates the same logical state as 
before, the additional folds of gates increase the number of operations by 
a factor of $2n+1$. This folding is often referred to as the \textit{global 
folding}.

One can also fold the individual gates ($g$) involved in the construction of 
$U$. This leads to the transformation of the native circuit
\begin{equation}
    U = G_gG_{g-1}...G_1
\end{equation}

to a new circuit $U^{'}$ described as
\begin{equation}\label{local_folding}
    U^{'} = G_g(G_g^{\dagger}G_g)^nG_{g-1}(G_{g-1}^{\dagger}G_{g-1})^n...G_1(G_1^{\dagger}G_1)^n
\end{equation}

The depth or the number of gates of the transformed circuit is given as 
$g^{'} = (2n+1)g$, and thus the scale factor can be given as $\lambda = 
g^{'}/g = 2n+1$. This method is referred to as \textit{local unitary 
folding} and has the added advantage of being flexible in 
terms of the choice of gates where the folding is to be applied. Instead of 
folding all the gates as described in Eq. \eqref{local_folding}, partial 
folding can be performed where only specific gates are folded. The choice of 
gates to be included in the set can be selected in chronological
order, i.e., from left or right, or chosen randomly. Random 
selection promises to be more economical for dUCC ansatz as it properly 
scales up the noise by randomly sampling gates from the entire circuit 
structure without completely folding the entire deep circuit. An illustration of the random local folding for circuit utilized in the evaluation of Term 3 (Eq. \eqref{12}) for $H_2$ molecule on \textit{ibmq\_manila} is given in Fig. \ref{manila_scaled}. Once the diagonal expectation
values are measured at different scale factors, the choice of an optimum 
extrapolation method plays a crucial role in the efficacy of the ZNE-PQE process. In the next section, we briefly overview various extrapolation methods that can be employed in the ZNE-PQE framework for accurately determining the ground state energies.

\subsubsection{Extrapolation Methods}
Another vital aspect of ZNE-PQE is the extrapolating method used on the diagonal values obtained using scaled-up noise.
This method works on the philosophy that the measured values for scaled circuits, denoted as $D(\lambda)$, follow some functional dependency on the scale factor $\lambda$. To approximate this relationship, one can use a parameterized function model. This approach simplifies the problem of determining $D(\lambda=0)$ by converting it to a straightforward regression, where the optimized model parameters can be obtained using a least-squared error approach. In this study, we provide a brief overview of various non-adaptive and adaptive extrapolation models that can be employed in ZNE-PQE and test their efficacy in the section on Results and Discussions.

\paragraph{Richardson Extrapolation} In this method, the dependence between the diagonal expectation values and the noise scale factor is considered to be $R$ degree polynomial in $\lambda$, where $R+1$ is the number of distinct noise scale values.  
The model can be given as follows: 
\begin{equation}
    D(\lambda) = c_0 + \sum_{i=1}^{R}c_i\lambda^i
\end{equation}
The zero noise value of D i.e. $D(0)$ can either be determined by solving $R+1$ linear equations using the interpolating \textit{Lagrange polynomial},
\begin{equation}
    D(0) = c_0 = \sum_{i=0}^{R}D(\lambda_i) \prod_{i\ne j} \frac{\lambda_j}{\lambda_j - \lambda_i},
\end{equation}

or can be explored using the least square fit technique. An important property of this method is that the uncertainty in estimating the zero noise value increases exponentially as the number of noise scale nodes increases.\cite{giurgica2020digital}

\paragraph{Adaptive Exponential Extrapolation} This model assumes the exponential dependence between the diagonal values and the noise scale factors. It can be represented as:
\begin{equation}\label{exponential_model}
    D(\lambda) = c_0 + c_1e^{-c_2\lambda}
\end{equation}
Unlike the previous model, where the noise scale factors are increased monotonically, this method chooses new $\lambda$ values adaptively based on the knowledge of the previously collected expectation values.
This model assumes that the value \enquote{$c_0$} is known for the asymptote($\lambda\rightarrow\infty$).  Thus, the optimization essentially boils down to finding the optimized value of $c_2$ (and $c_1$). The expression used is:
\begin{equation}
    c_2(\lambda_{j} - \lambda_{j+1}) = \alpha, \quad j \ge 0.
\end{equation}
where, $\alpha$ = 1.27846 is a constant. The adaptive process starts by setting the value of $j$ to $0$ and choosing an initial guess for the variable $c_2$, let's say $c_2 = 1$. Using that, $\lambda_{j+1}$ is calculated, which becomes the new noise node at which a new expectation value is measured. Using the best fit with these diagonal values in the exponential model, a new $c_2$ (and $c_1$) value is generated, which is then used to calculate the next $\lambda$ value. This process continues for a predefined number of steps, at the end of which, setting $\lambda=0$, the zero noise value is obtained. One can, of course, use the same model without the adaptive determination of the scale factors by determining all the model parameters ($c_0$, $c_1$, and $c_2$) through least squares fit on the predefined $\lambda's$.

\paragraph{Linear Extrapolation}Linear Extrapolation is one of the simplest methods for extrapolation. The relationship between the diagonal values and the noise scale factor is assumed to be linear. The model can be represented as
\begin{equation}
    D(\lambda) = c_0 + c_1\lambda
\end{equation}
where $c_0$ is the zero noise limit of the diagonal value. The optimized model parameters, $c_0$ and $c_1$, can be determined through regression.

In the Results and Discussion section, we present an in-depth analysis of ZNE-PQE with different choices of extrapolating models. This will help gain insight into the pros and cons of each method and establishes the optimum setting that incurs maximum accuracy. This section also dissects the residue norm landscape for the ZNE-PQE with the objective of identifying the underlying reason associated with the enhanced performance of ZNE-PQE. This comprehensive analysis will subjugate the behavior of ZNE, specifically in the PQE framework.
\begin{figure*}[!ht]
    \centering
\includegraphics[width=0.9\textwidth]{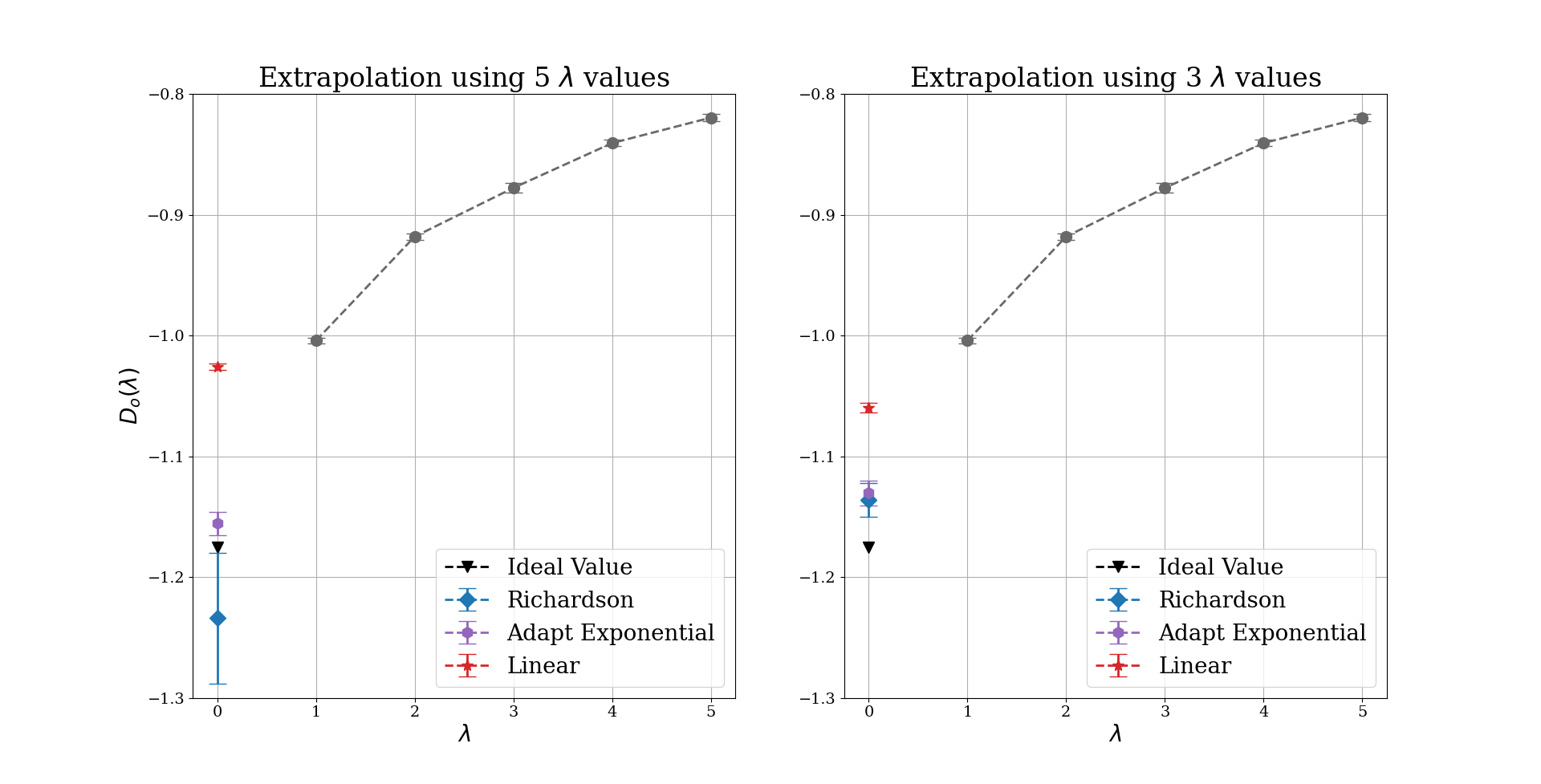}
\caption{Zero noise extrapolation of Term 3 for $H_2$ using noise model derived from \textit{ibmq\_manila}. In the left plot, extrapolation is done over $\lambda=\{1,2,3, 4, 5\}$ whereas in the right plot, it is done using $\lambda=\{1,2,3\}$. The asymptote for the Adapt Exponential is chosen to be $-0.8$. The entire process of scaling and extrapolating is repeated fifty times. The markers represent the average values obtained at different noise levels, whereas the error bars represent the standard deviation.}
    \label{trajectory}
\end{figure*}

\section{Results and Discussion}

\begin{figure*}[!ht]
    \centering
\includegraphics[width=\textwidth]{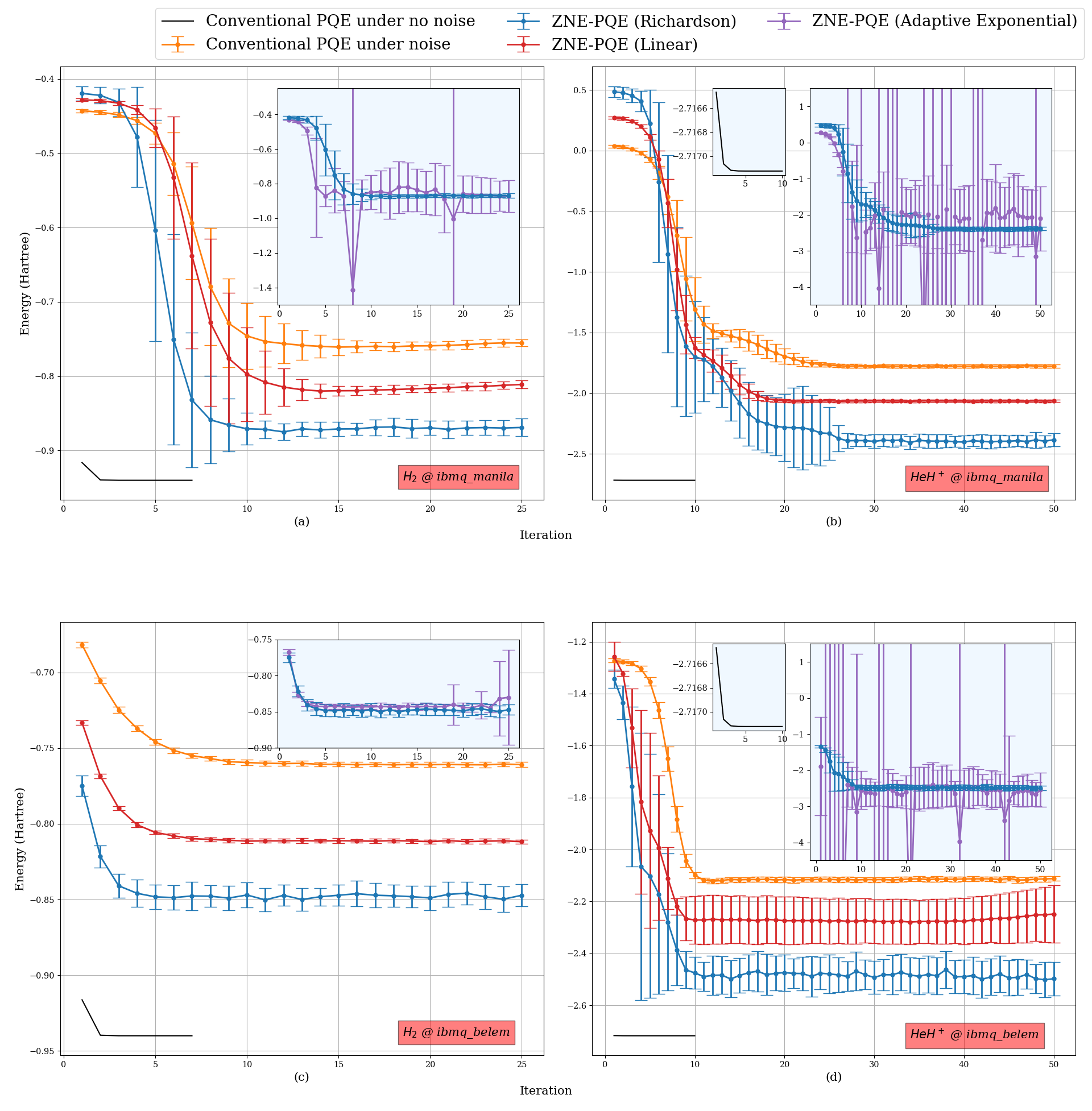}
\caption{Energy Vs. Iteration plot for $H_2$ ($r_{H-H}=2.25$\AA)  and $HeH^+$ ($r_{He-H}=0.55$\AA) with noise model and coupling derived from \textit{ibmq\_manila} and \textit{ibmq\_belem}. The noiseless and unmitigated are for conventional PQE. Whereas other curves represent ZNE-PQE with the respective extrapolation method. Each trajectory is obtained by averaging data from fifty runs with the standard deviation represented by error bars. For $H_2$, the inset provides the comparison between Richardson and Adapt Exponential. In the case of $HeH^+$, the insets provide the aforementioned comparison along with a zoomed plot of the noiseless conventional PQE trajectory (top left inset).}
    \label{energy_trajectory}
\end{figure*}

In this section, we begin by studying the efficiency of different extrapolation models by comparing their predicted zero noise extrapolated values(D(0)) with the corresponding ideal value ($Tr(H\rho)$). This provides us with an initial understanding of the factors affecting these models' performance and their viability for use in the ZNE-PQE framework. Using two different backends\cite{Qiskit}, we then analyze the energy convergence trajectory during the ZNE-PQE iterative optimization scheme for two molecules: $H_2$ and $HeH^+$. To construct these backends, we incorporate the shot-based \textit{qasm\_simulator} \cite{Qiskit} with noise models imported from IBM's \textit{ibmq\_manila} and \textit{ibmq\_belem} devices. These noise models include:
\begin{enumerate}
    \item Single qubit gate errors composed of \textit{depolarizing error} channel followed by the \textit{thermal relaxation error} channel on the respective qubit.
    \item Two qubit gate errors composed of two-qubit \textit{depolarizing error} channel followed by single qubit \textit{thermal relaxation error} on the involved qubits.
    \item Single qubit readout-errors on measurement.
\end{enumerate}

Finally, we perform a comparative study of the residue landscape to gain insight into the underlying reasons for the characteristic trajectory of ZNE-PQE and compare it with the conventional PQE (under noise as well as without noise). All the circuits are transpiled according to the basis gates the respective quantum devices support. The required integrals are obtained from pySCF\cite{sun2018pyscf} after Restricted HF calculations using an STO-3G basis set. We use the Jordan-Wigner transformation to obtain the components in terms of Pauli gates, and all implementations are performed using available qiskit-nature modules\cite{Qiskit}. All the ZNE-PQE-based calculations were performed by employing Mitiq\cite{larose2022mitiq} with an interface with qiskit.

\subsection{Insight Into the Efficiency of Different Extrapolating Models}

To obtain a general intuition about the efficiency of various models used during the extrapolation step, we plot the values of $D_o(\lambda)$ at different noise levels (simulated by random local unitary folding) along with the zero noise values obtained after extrapolation using three different models viz Richardson, Adapt Exponential and Richardson(Fig. \ref{trajectory}).

The comparison of different extrapolation models in terms of reproducing the ideal value reveals that the linear model performs the worst. It can be intuitively ascertained as the data values do not exhibit a linear dependency. The \textit{Adapt Exponential} and \textit{Richardson} models have comparative efficiency, with the former performing better when fitting is done over more points and the latter exhibiting lower variance and higher accuracy when fitted over fewer data points. The Richardson model approximates the diagonal values' dependency on $\lambda$ as a polynomial of degree $R$, where $R+1$ is the number of points used for extrapolation. The uncertainty increases exponentially with an increase in $R$, as evident from Fig. \ref{trajectory}. However, this analysis only provides an intuitive idea about the models' performance; their practical suitability for the ZNE-PQE framework will be further tested in the following section III B.

\subsection{Comparative Analysis of Different Extrapolating Models Towards ZNE-PQE Iterative Trajectory}

In this section, we study the iterative pathway as prescribed by ZNE-PQE in the presence of noise.
For this purpose, we have taken two systems viz $H_2$ ($r_{H-H}=2.25$\AA) and $HeH^+$ ($r_{He-
H}=0.55$\AA) and run the iterative ZNE-PQE method with dUCCSD ansatz in noisy simulators (Fig \ref{energy_trajectory}). These simulators have been 
prepared, as described previously, to mimic the predominant noise channels 
associated with real quantum devices viz \textit{ibmq\_manila} and 
\textit{ibmq\_belem}. The augmented noise is introduced through \textit{random local folding}. Three different extrapolation techniques - \textit{Linear}, \textit{Richardson}, and \textit{Adapt Exponential} have been explored for each molecule in each simulator.

For \textit{linear} and \textit{Richardson} models, the 
extrapolation is done on three points viz $\lambda=[1,2,3]$. For 
\textit{adaptive exponential} method, the number of scale factors was 
limited to five with the asymptote chosen to be -0.8 for $H_2$ and -3.0 for $HeH^+$. These values for the asymptotes require a preliminary analysis as done in Fig. \ref{trajectory}. It may be noted that the circuits for the augmented noise are assumed to be attainable in the NISQ devices. One can obviously restrict the maximum scale factor according to the quantum hardware at hand and choose the noise nodes accordingly. The parameters for Linear and Richardson models were 
determined through the least squares fit method, whereas for Adapt exponential, they were determined as outlined in section II B (2). The diagonal
values at each $\lambda$ are obtained for all cases by averaging over 5 measurements. The 
number of shots used for measurement in each experiment was fixed at 8192. The energy trajectory of the ZNE-PQE framework has been compared with the conventional PQE run at the identical noisy backend. We also provide the ideal PQE trajectory without any noise.

As discernible from these plots, ZNE-PQE provides substantial accuracy gain throughout the iterative trajectory as compared to conventional PQE under the noise. As expected, the Linear model performs the worst as compared to other models. While the Adapt Exponential provides, in general, similar performance as that of the Richardson, the high standard deviation and compromised stability in the former make it a less favorable method of choice for the ZNE-PQE framework. Richardson Model excels ahead of the rest, both in terms of accuracy and stability throughout the iterative trajectory. An important point to note is that different models' performance varies depending on the noise signature of different quantum devices, as can be verified from Fig. \ref{energy_trajectory}. But nevertheless, the relative performance of different models remains the same irrespective of the noise characteristics or the molecular system under study. This leads us to the conclusion that Richardson extrapolation is the most suitable candidate for the ZNE-PQE, which can outperform other models in terms of accuracy and stability on any given quantum device.

\subsection{Study of Residue Norm Landscape}

While the plots presented in the previous section describe the signature 
energy trajectory of ZNE-PQE for each molecule under a specific noisy backend, it is imperative 
to study the residue norm landscape for ZNE-PQE and how it varies from that of the conventional PQE. This will help in explaining the shift that is observed going from conventional PQE to ZNE-PQE. 
 \begin{figure}[!ht]
\centering
\includegraphics[width=\linewidth]{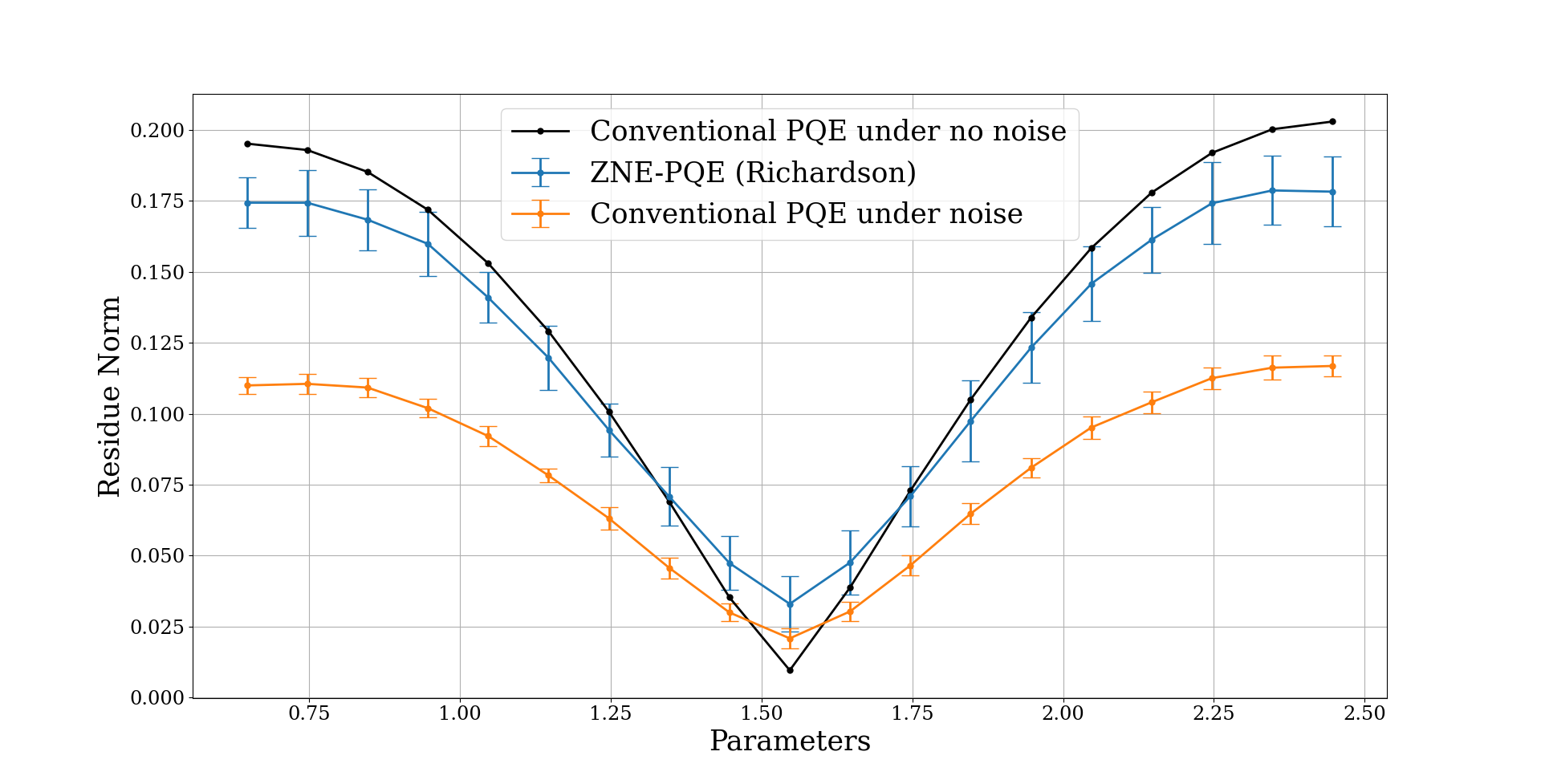}
\caption{Residue Norm ($\|r_{\mu}\|$) Vs Parameters plot for $H_2$ ($r_{H-H}=2.25$\AA) on a backend with noise model derived from \textit{ibmq\_manila}. Each point is the average of fifty residue evaluations, with the error bar denoting the standard deviation. The variation of the first ansatz parameter is captured in the x-axis, whereas the remaining ones are kept fixed.  The noiseless and unmitigated plots are for conventional PQE.}
    \label{norm_residue}
\end{figure}
We provide a residue norm vs. parameters plot in Fig. \ref{norm_residue} for ZNE-PQE (with Richardson model) and compare it with the conventional PQE for $H_2$. We also provide the noiseless PQE data for reference.
In general, a higher norm 
away from the fixed point indicates longer iterative strides resulting in 
a steeper decrease in energy. From Fig. \ref{norm_residue}, it can be observed that the residue norm for ZNE-PQE has higher values compared to the conventional PQE for parameters away from the point where the norm is minimum. This translates to a steeper decrease in energy, which can be corroborated by Fig. \ref{energy_trajectory}.

\section{Conclusions and Future Outlook}
In this study, we have developed the optimal formalism for the ZNE-PQE protocol to evaluate the ground state energies of molecular systems. Specifically, we parameterized the diagonal terms obtained from a quantum device using the noise scaling parameter and extrapolated them to their zero noise limit. This protocol has enabled us to improve the accuracy and efficiency of PQE significantly.

To test the effectiveness of different extrapolating models, we have extensively analyzed the energy trajectory obtained during the iterative optimization of the ansatz parameters in the presence of noise models from two different real quantum devices: \textit{ibmq\_manila} and \textit{ibmq\_belem}. The noise models we used consisted of the depolarizing error channel, thermal relaxation channel, and readout errors that typically occur during measurements.

Our analysis has provided insights into the efficacy of various extrapolating techniques and identified the optimal ZNE-PQE components that produce the most accurate and stable results. Additionally, we have analyzed the residue norm landscape for ZNE-PQE and compared it with the conventional PQE under identical noise and with the PQE under no noise. This comparison has helped us to understand why steeper energy trajectories occur in the case of ZNE-PQE. Overall, our study highlights the potential benefits of employing ZNE-PQE for evaluating ground state energies of molecular systems and provides valuable insights into the best practices for optimizing the performance of this approach in the presence of hardware noise. 

In this study, we used ZNE-PQE to mitigate generic gate errors effectively. While other mitigation techniques can also be compounded to reduce errors from less dominant sources, we have focused solely on ZNE. However, readout error mitigation techniques can be applied on top of ZNE to enhance the accuracy of ZNE-PQE further. Future research can be devoted to exploring the effectiveness of using multiple error mitigation techniques simultaneously in the PQE framework.

\section{Supplementary Material}
See the supplementary material for the characteristic properties and connectivities of IBM's superconducting quantum devices - \textit{ibmq\_manila} and \textit{ibmq\_belem}.

\section{Acknowledgments}
SH acknowledges the Council of Scientific \& Industrial Research (CSIR) for their fellowship. RM acknowledges the financial support from Industrial Research and
Consultancy Centre, IIT Bombay, and
Science and Engineering Research Board, Government
of India.

\section*{AUTHOR DECLARATIONS}
\subsection*{Conflict of Interest:}
The authors have no conflict of interest to disclose.

\section*{Data Availability}
The data is available upon reasonable request to the corresponding author.

\section*{References:}

%

\end{document}